\documentclass[12pt,preprint]{aastex}

\def\msol{M$_{\odot}$}
\usepackage{latexsym,amsmath,amssymb}
\usepackage{natbib}
\usepackage{rotating}
\def \degmark{$^\circ$}
\def \nh {N${\rm _H}$}
\def \ergsec{erg s$^{-1}$}

\def \chisq {$\chi ^{2}$}
\def \rchisq {$\chi_{\nu} ^{2}$}
\def \cha {$Chandra$}
\def \swi {$Swift$}
\def \xmm {$XMM$-$Newton$}

\shorttitle{ 
{\it SWIFT} Observations of Nova-likes  
}
\shortauthors{Balman et al.}

\begin{document}

\title{
{\swi } XRT Observations of the Nova-like Cataclysmic Variables MV Lyr, BZ Cam and V592 Cas  
} 

\author{\c{S}\"olen Balman} 
\affil{Department of Physics, 
Middle East Technical University,
Ankara, Turkey}
\email{solen@astroa.physics.metu.edu.tr}   

\and

\author{Patrick Godon\altaffilmark{1}, Edward M. Sion } 
\affil{Astronomy \& Astrophysics, Villanova University, \\ 
800 Lancaster Avenue, Villanova, PA 19085, USA}
\email{patrick.godon@villanova.edu, edward.sion@villanova.edu} 

\altaffiltext{1}{Visiting in the Henry A. 
Rowland Department of Physics and Astronomy, 
The Johns Hopkins University,
Baltimore, MD 21218. 
}

\begin{abstract} 
 
We present a total of $\sim$45 ksec (3$\times$15 ksec) of \swi\ XRT observations for  
three non-magnetic nova-like (NL) Cataclysmic Variables (CVs) (MV Lyr, BZ Cam, V592 Cas)
in order to study characteristics of Boundary Layers (BL) in CVs. 
The nonmagnetic NLs are found
mostly in a state of high mass accretion rate ($\ge$1$\times$10$^{-9}$ \msol\
yr$^{-1}$) and some show occasional low states. Using the XRT data,
we find optically thin multiple-temperature cooling flow
type emission spectra with X-ray temperatures (kT$_{max}$) of 21-50 keV. These hard X-ray
emitting boundary layers diverge from simple isobaric cooling flows indicating
X-ray temperatures that are of virial values in the disk. In addition, we
detect power law emission components
from MV Lyr and BZ Cam and plausibly from V592 Cas which may be a result of
the Compton scattering of the optically thin emission
from the fast wind outflows in these systems and/or
Compton up-scattering of the soft disk photons. The X-ray luminosities of the 
(multi-temperature) thermal plasma emission
in the 0.1-50.0 keV range are (0.9-5.0)$\times$10$^{32}$ erg/sec. The ratio of the
X-ray and disk luminosities (calculated from the UV-optical wavelengths) yield an efficiency
(L$_{x}$/L$_{disk}$)$\sim$0.01-0.001. Given
this non-radiative ratio for the X-ray emitting boundary layers with no
significant optically thick blackbody emission in the soft X-rays (consistent
with $ROSAT$ observations) together with the high/virial X-ray temperatures,
we suggest that high state NL systems  may have optically
thin BLs merged with ADAF-like flows and/or X-ray coronae. In addition, we note that the
axisymmetric bipolar and/or rotation dominated fast wind outflows detected in
these three NLs (particularly BZ Cam and V592 Cas) or some other NL may also
be explained in the context of ADAF-like BL regions.

\end{abstract} 


\section{Introduction} 

Cataclysmic Variables (CVs) are short period (up to $\sim$2 day) 
close binary systems in which
a white dwarf (WD) accretes matter from a late-type main sequence star 
filling its Roche lobe (Warner 1995). 
In non-(or weakly-) magnetic CVs the transferred material 
forms an accretion disk around the WD  
and reaches  all the way to
the stellar surface, as the magnetic field is not strong enough to disrupt the 
accretion disk.

NL systems are a class of CVs usually found in a state of high mass accretion 
rate, though some NLs are sometimes found in a low state 
of reduced accretion. NLs have two main subclasses where VY Scl stars exibit
high states and occasional low states of optical brightness and the UX UMa stars
remain in the high state and low states are not seen (Warner 1995).
RW Tri stars are NLs that are eclipsing UX UMa systems (so they're all high inclination NLs).    
While all NL variables show emission lines, UX UMa sub-type of 
NLs also exhibit broad absorption lines in the optical and/or UV wavelengths.  
In the high states of NLs, the accretion rates 
are typically a few $\times$10$^{-8}$\msol\ yr$^{-1}$ to a few $\times$10$^{-9}$ \msol\
yr$^{-1}$. Virtually all NLs reside above the period gap with a concentration of
them between P$_{orb}$ of 3 and 4 hours (Ritter \& Kolb 1998). There is only one NL 
below the period gap (BK Lyn), this system has been
identified with a nova in ancient times from chinese records (see Patterson et al. 2013).
The high accretion rates are inferred from luminous disks 
seen best and modeled in the FUV with steady-state disk models.
In addition, bipolar outflows and/or rotationally dominated winds from NLs are 
detected primarily in the FUV, and P Cygni profiles of the resonance doublet of 
CIV are
almost invariably seen from these collimated outflows in 
systems with disk inclinations less than 60-70 degrees 
(Guinan \& Sion 1982, Sion 1985).
The high accretion rates may be driven in part by irradiation
of the donor star if the accreting WD remains hot (e.g., long 
after a nova event). Existence of hot WDs in NLs in turn leads to irradiation of the 
donor hence a driving of the high mass transfer by the donor, either 
by irradiation driving a wind off of the donor or by irradiation causing the Roche-Lobe filling 
donor star to be bloated. There is another subclass of nonmagnetic CVs called SW Sex stars often
listed as part of NLs. These are defined by specific spectroscopic characteristics and have orbital
periods between 3-4 hrs (see Rodriguez-Gil et al. 2007 for a review).
Not all NLs show these characterstics. Our present paper does not deal with SW Sex stars.

Observations of CV disk systems at low mass accretion rate 
(namely dwarf nova CVs in quiescence) have successfully led to the  
determination of the temperature and luminosity 
of the accreting white dwarfs from FUV observations (where it is the dominant
emitting component at low $\dot{\rm M}$) and X-ray observations
have yielded  the temperature and luminosity of 
the optically thin BLs.   
The dwarf novae in quiescence (low $\dot{\rm M}$ systems) were observed with
recent X-ray telescopes like \xmm, \cha, and $Suzaku$\
(e.g. Szkody et al. 2002, Pandel et al. 2005,
Rana et al. 2006, Okada et al. 2008, Ishida et al. 2009, Mukai et al. 2009, Balman et al. 2011) 
and the results seem in accordance with one-dimensional 
numerical simulations of the 
optically thin BLs in standard steady-state disks of DNe in quiescence with  
low accretion rates (10$^{-10}$-10$^{-12}$ \msol\ yr$^{-1}$, (Narayan \& Popham 1993, Pringle 1981).
However, note that a recent study of X-ray variability and inner disk structure
of DN in quiescence  
reveals optically thick  disk truncation and plausible formation of hot (coronal) flows
in the inner parts of the quiescent DN accretion disks (Balman \& Revnivtsev 2012).

One of the earliest comprehensive
studies on hard X-ray emission from the boundary layers of CVs using the $Einstein$ IPC (0.2-4 keV) 
indicate that NL systems (5 detected systems) emit hard X-ray emission in this range with 
luminosities $\le$ a few$\times$10$^{32}$ \ergsec (Patterson \& Raymond 1985).  
NL systems have been
detected in the early epochs of $ROSAT$ Observatory (about 11 NL systems; 0.1-2.4 keV) and
optically thick soft X-ray components expected from these high accretion rate systems are not detected 
(van Teeseling et al. 1996). For these systems a hot optically thin X-ray 
source is found with plasma temperatures kT$<$ 6 keV. The X-ray luminosities are $<$ a 
few$\times$10$^{32}$ \ergsec\ whereas the UV luminosities are in a range
10$^{31}$-10$^{35}$ \ergsec.
In addition, Greiner (1998) has done a comprehensive analysis of the 
$ROSAT$ detected NLs (see also Schlegel \& Singh 1995) and have argued 
that $ROSAT$ data are also consistent
with blackbody models yielding lower \rchisq\ than thermal plasma models. 
These blackbody temperatures are calculated as (0.25-0.5) keV
with very small emitting regions.
Later, some NLs were studied with $ASCA$ and are found consistent with double MEKAL models
at different temperatures (e.g., TT Ari and KT Aur, Mauche \& Mukai 2002), and 
one with \xmm\ using multiple-temperature plasma and MEKAL models (Pratt et al. 2004).
In all these observations the X-ray luminosities are $\le$ afew $\times$10$^{32}$ \ergsec.

Recent advances in UV spectroscopy have revealed that the high mass accretion 
disks in nova-likes are departing
from the standard disk models especially in the inner disk (Puebla et al. 
2007), where the boundary layer is, also, located. 
Puebla et al. (2007) have done a comprehensive UV modeling of accretion disks
at high accretion rates in 33 CVs including many NLs and old novae.
The UV findings indicate
a necessity for improvement by incorporating a component from an extended optically
thin region (e.g., wind, corona/chromosphere). This is evident from
the strong emission lines and the P Cygni profiles observed in the
UV spectra.  Disk irradiation by the boundary
layer or the central star, and nonstandard temperature profiles would
help for better modeling in this band. An additional possibility to explain model
discrepancies is disk truncation. Yet another possibility is
to modify the viscosity law for example by using constraints on the viscosity
following from numerical simulations of the magneto-rotational instability. 
A first step in improving the disk modeling
has been to modify the inner disk to reflect the presence of a 
hot boundary layer (e.g. for QU Car: Linnell et al. 2008;
for MV Lyr: Godon \& Sion 2011) where an additional hot optically thick component
was also found consistent with the data when modeling in the UV. 

\section{The Nova-like Systems MV Lyrae, BZ Camelopardalis, and
V592 Casiopeiae} 

In this paper, we present the \swi\ XRT observations of three NLs, MV Lyr, BZ Cam, and V592 Cas.
The important characteristics of the systems are presented in Table 1 and the
relevant literature is summerized below.

\subsection{MV Lyr} 

MV Lyrae is a member of  VY Scl subclass of NLs, 
mostly found in a 
a high state of accretion (similar to the outburst state of DNs but lasting much longer)  
with occasional drops (short-duration) in brightness. 
In these low states,
the magnitude of MV Lyr drops from $V\approx 12-13$ to $V\approx 16-18$ 
\citep{hoa04}. The archival AAVSO data reveals that it 
stays in a high state for up to $\sim 5$yr, and following that for a period
of a few years to $\sim10$yr it starts alternating between
high state and low state on a time scale of a few months to a year.  
MV Lyrae has an orbital period of  
$P=3.19$hr and a mass ratio of 
$q = M_{2nd}/M_{wd} =0.4$ \citep{sch81,ski95}. 
The inclination of the system is found in a range
$i=10^{\circ}-13^{\circ}$, constrained with the small radial velocity measurements
and lack of eclipses \citep{lin05}.

The UV observations of MV Lyr obtained with the 
{\it International Ultraviolet Explorer ({\it IUE})} during its low state
revealed the existence of a hot WD reaching 50,000K \citep{szk82} or 
higher \citep{chi82}.
Later, the source was observed in a different low state 
with {\it FUSE} (in 2002) that lasted about 8 months.  
The analysis showed that the WD has
a temperature of 47,000K, a gravity of $\log{g}=8.25$, a projected rotational
velocity of $V_{rot} \sin{i} =200$ km s$^{-1}$, sub-solar abundances of
$Z = 0.3 \times Z_{\odot}$, and a distance of $505 \pm 50$pc (see \citet{hoa04}).
In this low state, the magnitude of MV Lyr was around
$V\approx 18$ with a mass accretion rate about              
$\dot{M} \approx 3 \times 10^{-13}M_{\odot}$\ yr$^{-1}$ \citep{hoa04}.  
Furthermore, \citet{lin05} studied the different states of MV Lyr and its secondary
with the aid of {\it IUE} and HST/{\it STIS} spectra, 
and found that the  mass accretion rate was of the order of
of $3 \times 10^{-9}M_{\odot}$\ yr$^{-1}$
during the high state. 
\citep{god11} analyzed the  FUSE spectrum of MV Lyr in its high state
and found an accretion rate of about $2 \times 10^{-9}M_{\odot}$\ yr$^{-1}$,    
assuming $i=10^{\circ} \pm 3^{\circ}$ \citep{sch81,ski95,lin05},   
a 50,000K WD and an extended 100,000K UV emitting region which can 
originate from an optically thick BL or it can be from an inner disk region 
irradiated by the boundary layer (i.e., optically thin BL).

\subsection{The Peculiar Nova-like BZ Cam} 

BZ Cam is a peculiar NL with rather special characteristics 
among cataclysmic variables. First, it is associated with a 
bow-shock nebula \citep{ell84,kra87,gre01}; second, as a wind-emitting system
it manifests its outflow not only in the FUV resonance lines
(as do wind-emitting CVs) but also in the Balmer and He\,{\sc i}
lines \citep{pat96}; third, the wind shows a bipolar nature
as a highly unsteady and continuously variable $\sim$ 3000 km s$^{-1}$ supersonic outflow
 \citep{hon13}; last, the variability timescales seen in 
its optical wind-outflow features are much shorter
(minutes to hours) than the shortest variability 
reported in the FUV of any other CV wind \citep{pat96}.

For these reasons, BZ Cam has been the subject of many studies of
the time-variability of its line profiles
(e.g. \citep{kra87,hol92,woo92,gri95,pri00,hon13}). 
It is classified as a VY Scl type 
NL system with a period of 221 min
\citep{tho93,pat96}. 
\citet{gar88,gre01} present an analysis of its two optical low states.
The source has an inclination of about $i=12^{\circ}-40^{\circ}$ \citep{lad91,rin98}
and a distance of $830\pm160$pc. 

Its WD mass is unknown but it is thought
to contain a $0.3-0.4M_{\odot}$ main-sequence donor star \citep{luh85}.   
It is mostly seen at a magnitude of $V=12.0-12.5$ mag and during low
state reaches $V=14.3$ mag \citep{gre01}. It was observed with ROSAT
\citep{van96}, IUE \citep{kra87,woo90,woo92,gri95}, HST/GHRS \citep{pri00}
and FUSE \citep{fro12}. The UV and optical continuum is consistent with a
$\sim$12,500 K Kurucz LTE model atmosphere \citep{pri00} with a small emitting area 
compared to the surface of a WD. In work presented elsewhere 
(Godon et al. 2014, in preperation), 
our analysis of the archival FUSE spectra of
BZ Cam reveals a 50,000 K WD. 

\subsection{V592 Cas} 

V592 Cas is an UX UMa subtype of Nova-like with an inclination of $i=28^{\circ}$$\pm$10$^{\circ}$
\citep{hub98}, and a period of 2.76 hr \citep{tay98}.  With a reddening of 
E(B-V)=0.22 \citep{car89}, it has been placed at about 330-360pc \citep{hub98,hoa09}. 
Its WD is about 0.75 M$_{\odot}$ with a temperature of 45,000K \citep{hoa09}
and a mass accretion rate around 
1$\times$10$^{-8}$ M$_{\odot}$ yr$^{-1}$ \citep{tay98,hoa09}. 
V592 Cas has a bipolar wind outflow
that has an episodic nature, with several events 
reaching velocities of 
5000 km s$^{-1}$ in H$\alpha$
where the optical brightness variations and the  strength of the outflow 
reveals no clear correspondance (Kafka et al. 2009).
V592 Cas was observed with IUE \citep{tay98} and FUSE \citep{pri04} revealing
Doppler shifts of the entire blueward absorption troughs  
in the ultraviolet resonance lines as the main source of variability characterized with
an asymmetric and  non-sinusoidal behaviour over the orbital phase.
Furthermore, the outflowing wind does not show modulated variation
on the superhump periods detected from V592 Cas, but show modulation only on the orbital phase (orbital period).
This indicates that neither the precession of the disk nor the precession of the disk tilt (negative and positive superhumps)
is affecting the outflow.
In addition, V592 Cas reveals a short photometric oscillation of 22 mins \citep{kat02}.  

\section{The Standard Star-Disk Boundary Layer} 
 
The standard disk theory \citep{sha73,lyn74},  predicts that  
viscous dissipation of energy 
is instantly radiated locally in the vertical z-direction. 
The disk total luminosity $L_{disk}$ is half the accretion luminosity  
\begin{equation} 
L_{disk}=\frac{L_{acc}}{2} = \frac{G M_{*} {\dot{M}}} {2R_{*}},  
\end{equation} 
where $G$ is the gravitational constant, $M_*$ is the mass of the 
accreting star, $R_*$ its radius and $\dot{M}$ is the mass accretion
rate. 
Each face of the disk radiates $L_{disk}/2$.   

If the star is not rotating close to breakup
(i.e. $\Omega_{*} << \Omega_K(R_{*})$r), 
then the theory predicts (e.g. \citet{pri81}) that
the remaining rotational kinetic energy of the
material at the inner edge of the disk (rotating at the local
Keplerian speed) is dissipated in a small region as this material
lands on the WD which is rotating more slowly at sub-Keplerian speeds.
This region is refered as the boundary layer (BL) and up to half
the accretion luminosity is liberated in this region.  
This remaining rotational kinetic energy given-off from the BL
($L_{BL}$) is 
\citep{klu87}: 
\begin{equation}
L_{BL}= L_{disk}  \left(
1 - \frac{\Omega_{*}}{\Omega_K(R_{*})} \right)^2. 
\end{equation}
The BL luminosity is not reduced more than a factor of 1/4 due to rotational effects.
At high mass accretion rates ($\dot{\rm M} \approx  10^{-9}-10^{-8}$ M$_{\odot}$
yr$^{-1}$) the BL is largely expected to be optically thick and emits in the soft X-ray band 
\citep{pri77,pri79,pop95,god95,hert13}, 
and at lower mass accretion rates it expected to be
optically thin and emit in the hard X-ray ($\sim$ 10 keV) band (Narayan \& Popham 1993, Popham 1999). 
Furthermore, two-dimensional semi-analytical
work  \citep{pir04} shows  
that effects of the centrifugal force and pressure may result in,
rings of enhanced brightness above
and below the WD equator, called the ``spreading layer'' 
\citep{ino99}.  At lower mass accretion rates than
$ < 1.6\times10^{-8}$ M$_{\odot}$ yr$^{-1}$, however, \citet{pir04} 
shows that the spreading is negligible which is the case for the
three NL sources in this paper (see Table 1 for the accretion rates).  
The transition between an optically thin and an optically thick
boundary layer does not only depend on the mass accretion rate,
it also depends on the mass of the white dwarf, on its rotational velocity, and on the alpha
viscosity parameter (disk viscosity $\nu$= $\alpha$c$_s$H\ where c$_s$ is the sound speed,
H is the disk height and $\alpha$ is a free paramter in a range 0-1). 

\section{Observations and Data}

\swi\ was launched in November 2004 \citep{geh04} mainly designed to measure
position, spectrum and brightness of 
gamma-ray bursts (GRBs). 
It has a Burst Alert Telescope
(BAT; \citet{bart04}) with typically 100 sec reaction time and a field of
view (FOV) of 100\degmark$\times$60\degmark\  working in 15-150 keV range using a CdZnTe CCD.
The narrow field instruments are the X-ray telescope (XRT;
\citet{bur05}) and the Ultraviolet-Optical Telescope (UVOT; \citet{rom05}).
The XRT operates in the 0.2-10 keV range with an 18 arcsec half-power diameter at 1.5 keV
using CDDs that are similar to EPIC MOS on \xmm.
It has an FOV of 23.6$\times$23.6 arcmin and an imaging resolution of 2.4 arcsec per pixel.
Regular observations of MV Lyr, BZ Cam and V592 Cas were made with the \swi\ spacecraft 
between 2012 June 8 and 2012 December 21 utilizing both the XRT and the UVOT using the UV grism.
Table 2 gives the details and log of the observations.  
In this paper, we report on the analysis of the XRT data of the three NLs obtained in the 
photon counting (PC) mode for about 15 ksec each. Data were obtained, also, in the windowed timing
mode (WT) using short exposures less than 5 min, this mode does not provide spatial resolution.  
Such short exposure times will not yield adequate S/N for
any spectral or timing analysis, thus WT data have not been used for the analysis.
The UV grism analysis will be discussed in a subsequent paper. PC mode is a  
frame transfer operation of an X-ray CCD retaining full imaging and spectroscopic resolution, 
but the time resolution is only 2.5 sec; the mode is used at low fluxes below 1 mCrab. 
We find XRT count rates of 0.069(3) c s$^{-1}$ for MV Lyr, 0.051(2) c s$^{-1}$
for V592 Cas, and  0.070(3) c s$^{-1}$ for BZ Cam. We cross checked the states
of the sources using the existing AAVSO data and the UVW1 filter magnitudes and count rates,
obtained simultaneously with the PC mode data. We find that all the three NLs were in a high state
during the \swi\ observations.

The screened and pipeline-processed data (aspect-corrected, bias-subtracted, graded (0-12) and
gain-calibrated event lists) are used for the analysis. Latest calibration file on the \swi\
database (version, 20010101v013) is used, in accordance with the XRT data for the 
redistribution matrix. The ancillary response files are calculated for each source by
merging exposure maps in {\sc XIMAGE} created using $xrtexpomap$ task and finally running
the task $xrtmkarf$. For light curve and spectrum generation, and further analysis, 
XSELECT V2.4b and HEASoft 6.13 (see http://heasarc.nasa.gov/lheasoft/) are utilized. 

In addition, we have used $ROSAT$ archival data of our target sources to derive blackbody temperature upper
limits. The data for MV Lyr was obtained on 1992 November 4-8 using the PSPCB detector (20 ksec); OBSID rp300192n00.
For BZ Cam, the observation was performed on 1992 September 28-29 using the PSPCB detector (6.0 ksec); OBSID 
rp300233n00. Finally for V592 Cas, the data was obtained during the RASS (all sky survey; 1990 December 29 to 1991 August 06) 
using the PSPCC detector (0.6 ksec); OBSID rs930701n00. The three $ROSAT$ data sets were aquired in high states, note that
V592 Cas is a UX UMa type of NL which shows only high states.
For details of $ROSAT$ data see van Teeseling et al. (1996), Greiner (1998), and Voges et al. (1999).

\section{Temporal Analysis}

XRT light curves are extracted from the data sets of each NL sampling between 0.2-10 keV
at 2.7 sec resolution and background subtracted. A circular region of radius 
1.5 arc min is used for the photon extraction for both the source and the background
(devoid of other contaminating sources).
  
We searched for variations of the X-ray light curves  
over the orbit of the systems  
by folding each light curve on the orbital 
period using 10 phase bins. The results are displayed in Figure 1. 
For the orbital periods we used the spectroscopic periods derived for the three NLs:
(1) For BZ Cam, we used the period and the ephemeris derived from the He I $\lambda$5876 line
(T$_0$=2453654.008(2)$+$0.15353(4)$\times$E: Honeycutt et al. 2013); (2) For V592 Cas, we used
the ephemeris derived from the He I $\lambda$5876 and $\lambda$6678 line (T$_0$=2450707.866(1)$+$
0.115063(1)$\times$E: Taylor et al. 1998); (3) For MV Lyr, we used the ephemeris derived from the H$\alpha$
line (T$_0$=2449258.05$+$0.1329(4)$\times$E: Skillman et al. 1995). 
In Figure 1 each NL shows some variation of the X-ray count rates over the orbit of the system. 
V592 Cas shows 
complex behaviour, but looks in-phase with the H$\alpha$ variation (only the first narrow X-ray peak) 
and possibly out-of-phase with the He I $\lambda$5876 line.
Both in BZ Cam and V592 Cas, the He I $\lambda$5876 line and H$\alpha$ line variations 
are out-of-phase.
V592 Cas shows dips close to zero counts per sec and the X-ray variations seem more drastic 
compared to BZ Cam or MV Lyr. 
The second maximum of X-rays seem to be
right before/at around when the wind strength maximizes (over the orbital phase)  
as the H$\alpha$ line equivalent widths maximize
calculated from the blue shifted P Cygni absorption profiles. Note that for V592 Cas, both 
the He I $\lambda$5876 line and the H$\alpha$ line have a single peak variation over the orbit, whereas the 
X-rays show double peaks within the orbit relevant to the bipolar wind nature of the source. 
For BZ Cam, the maximum X-ray variation is about 67$\%$ in the X-ray count rate and the 
X-ray variation of MV Lyr is about 50$\%$ in the count rates. For these two sources,
the relative phasing between the X-rays and the optical is not definitive because the error
on the orbital periods fold close to one full phase since the given time zero. For
V592 Cas, erorr on the relative phasing between the optical and the X-rays are less than 0.05.   
We cannot conclusively calculate the modulation amplitudes or
significance of the X-ray  variations over the orbit since the data is about 15 ksec for each NL 
and covers only less than two cycles of 
the orbit at best. For MV Lyr, a complete coverage does not exist. 
We notice that the X-ray variations
for these sources do not follow a sinusoidal shape over the orbit. 
Analyses of energy dependence
of X-ray variations are performed using two energy bands of 0.3-2.2 keV 
(or 0.3-1.0 keV) and 2.5-7.0 keV revealing
no significant change in the shape of the mean light curves over these energy bands, 
given the statistical errors of the low count rates which indicates no sigificant energy 
dependence in variations over the \swi\ XRT range.
 
In addition, we looked for  other periodicities or coherent QPOs (quasi-periodic oscillation)
that could exist in the XRT data
using fast fourier transform (FFT) analysis and/or averaging several power spectra by segregating 
data into several smaller time intervals.
We found no significant periodicity or coherent QPO  
for BZ Cam, V592 Cas or MV Lyr given the S/N quality of the data. 
   
\section{Spectral Analysis}

A spectrum and a background spectrum 
was generated for each NL using all available data presented in Table 2 between 0.2-10 keV.
A circular photon extraction region of radius 
1.5 arc min is used for both the source and the background
(devoid of other contaminating sources). Each spectrum was grouped to have a minimum of 30-60
counts in each bin to increase signal to noise and utilize good  \chisq\ statistics.
We emphasize that the data is of moderate spectral resolution and no significant line emission
other than predicted by the fitted models was detected; particularly, no
iron lines at 6.4 keV, 6.7 keV or 6.9 keV was found, and so no reflection
component can be confirmed with this data. However, NLs have hot disks with winds, as opposed to cold disks of
quiescent DN, which may yield complicated reflection effects that may not be revealed properly with \swi\ data
in the 0.2-10.0 keV band.
Subsequently, these spectra are analyzed using single/double/triple thermal plasma emission models
MEKAL/APEC (thermal plasma in collisional equilibrium)  within XSPEC software or multi-temperature
isobaric cooling flow (plasma) 
models CEVMKL or MKCFLOW in XSPEC (for references and model descriptions see Arnaud 1996,
or https://heasarc.gsfc.nasa.gov/xanadu/xspec/manual/Models.html). 
To account for the
absorption in the X-ray spectrum of interstellar or possible intrinsic origin, we utilized
the $tbabs$ model in the fitting procedure (Wilms, Allen \& McCray 2000).  

In general, the X-ray spectra of nonmagnetic CVs are well modelled with the multi-temperature
isobaric cooling flow type plasma emission models as in MKCFLOW or CEVMKL (Mukai et al. 2003,
Baskill et al. 2005, Pandel et al. 2005, Guver et al. 2006, 
Okada et al. 2008, Balman et al. 2011, Balman 2014).
The X-ray spectra of nonmagnetic CVs show several different emission lines, revealing
the existence of  hot, optically thin plasma in these systems (see also Balman 2012).
A wide range of temperatures is derived from these spectra
revealed by the presence of detected H- and He-like emission lines from 
elements N to Fe, and some Fe L-shell lines. 
As the accreting material settles onto the WD through the BL region, it is expected
to form a structure with continuous temperature distribution in the X-rays.
This characteristic is consistent with an isobaric cooling flow plasma model 
which is a multi-temperature
distribution of plasma with a differential emission measure assuming a power-law distribution
of temperatures ($dEM=(T/T_{max})^{\alpha -1} 
dT/T_{max}$). In such a model, the
emission measure at each temperature is proportional to the
time the cooling gas remains at this temperature (Pandel et al. 2005).  This type of plasma emission 
represent collisionaly ionized plasma in thermal equilibrium between ions/protons 
and electrons just like
the standard MEKAL/APEC models assume.

The XRT spectrum of\ BZ Cam was fitted using several plasma models within XSPEC. A single  
or double MEKAL model fits yield unacceptable \rchisq\ values of 2.6 and 2.1, respectively.
Increasing MEKAL components reduces the \rchisq\ artificially,  
but approximates a multiple-temperature distribution plasma. 
Therefore, as discussed above, a more physical model CEVMKL which 
is a multi-temperature plasma emission model built from the $MEKAL$ code (Mewe et al. 1986) is used
to fit the data. However, the data was found inconsistent with a \rchisq\ of 2.7\ .
BZ Cam has strong wind outflows (in the UV and optical bands) 
as discussed in the introduction and it may be that
there is an excessive component in the X-ray spectrum due to scattering off of the wind
or an emission component due to an extended structure on the disk. 
We used a power law model to characterize this second component
and fitted the spectrum with $tbabs$$\times$(CEVMKL+POWER) composite model. The resulting
spectral parameters with an acceptable \rchisq\ are given in Table 3 (middle column) and
the fitted spectrum is displayed in Figure 2.
We carefully checked that all spectral parameter ranges are consistent across spectral
binning of 30-60 counts per bin for the fitted model with the  minimum \rchisq\ values. 
However, the \rchisq\ values of the unacceptable fits in this paragraph are even higher 
when the spectrum with 60 counts per bin is used which is expected since the
greater binning increases the discepancies between the model and data reducing statistical errors.
The excess data are in the harder energy band.
 
For fitting the XRT spectrum of MV Lyr, we assumed similar models. A single MEKAL model
fit yields an unacceptable \rchisq\ value of 2.64\ . A double MEKAL model gives acceptable fits
with a lower temperature of kT=0.2 keV and a higher temperature of kT=79.8 keV (a 2$\sigma$ lower limit
on the high temperature MEKAL model is kT=29 keV). This signals a definite distribution of temperatures
where the MEKAL components somehow reflect the lower and higher temperature limits. 
Thus, we fitted the 
spectrum with a CEVMKL model alone, which yielded a \rchisq\ value of 1.7(dof.21). In order to look for 
any scattering effect of X-rays from a wind or a component relating to an extended structure, 
as in BZ Cam (note that
this system is also known to have winds in the UV/optical), we added a power law component
and calculated how much this composite model $tbabs$$\times$(CEVMKL+POWER)
improves the fit performed using only the CEVMKL model. The spectral parameters and the \rchisq\ value of
this new/composite fit is given in Table 3 (left hand column), and the modeled spectrum is presented
in Figure 3 . We tested the significance of adding the
second power law component to the CEVMKL model with FTEST and found that it improves the fit
at 98.6\% confidence level (probab. 0.014), which is almost at 3$\sigma$ significance.
Note that when the binning in the spectrum is increased to a minimum of 60 counts per bin,
the single CEVMKL fit has a \rchisq\ value of 2.0 which is then improved over 3$\sigma$ significance
when the power law model is added to the fit. The power law model removes the excess data in the harder energy
band. Increasing the binning improves the S/N of the
spectral bins which in turn affects the \chisq statistics of spectral fits that 
do not model the spectrum adequately. This method reveals a correct spectral model that describes the 
spectrum,
given the moderate spectral resolution and S/N of the data.       

Our final target, V592 Cas, was the lowest count rate source of the three NLs.
We treated the modeling of the XRT spectrum in a similar manner as in the other two.
However, in this case it is more difficult to chose between the models relying only on
\rchisq\ values due to the lower statistical quality of the data. 
A single MEKAL model fit yields a  \rchisq\ value of 1.5 with an
X-ray temperature of kT=41 keV (a 2$\sigma$ lower limit is kT=15 keV). Adding a second MEKAL
component yields two temperatures of 0.4 keV and 8.1 keV improving the \chisq\ of the 
fit at 92\% confidence 
level (probab. 0.08 using FTEST). Inclusion of another MEKAL does improve the fit even more, 
but then it is redundant since it indicates existence of a distribution of temperatures.
Thus, for the consistency of the modeling of V592 Cas with the other two NLs, we included
a fit using the model $tbabs$$\times$CEVMKL yielding a  \rchisq\ value
better than a single MEKAL fit at 94\% confidence level (using FTEST). We present this
fit in Table 3 (right hand column) and the fitted specrum is shown in Figure 4.
In this case, inclusion of a secondary power law
component does not improve the fit (yields \rchisq\ value of 1.25). However,
we include maximum limits on the power law flux and photon index for completeness.
We caution that the non-detection of a significant power law component could be a result of the 
lower X-ray flux of the source. 

The unabsorbed X-ray flux and luminosities (calculated from the distances in Table 1) 
of the three NL systems are given in Table 3 along with the rest of the
spectral parameters with errors stated at the 90\% confidence level for a single parameter. 
We caution  that due to the energy range of \swi\ XRT and the
hardness of the X-ray spectra, we may have not determined
the power law photon indices as accurately and they
may be in a steeper range as in 1.0-2.0 as opposed to 0.5-1.0 as seen from Table 3.
Given the unabsorbed total X-ray fluxes of these sources (also, using the power law photon indices
and fluxes) and the XRT count rates,  \swi\ BAT is not expected to detect these sources, with an expected
BAT rate $\le$ 2$\times$10$^{-4}$ c s$^{-1}$. We have analyzed the BAT data and have not detected any
of the sources, as expected. 

The range of neutral hydrogen column densities derived from the spectral fits (see Table 3),
are checked using standard tools that calculate neutral hydrogen column in the line of sight:
(1) COLDEN (using Dickey \& Lockman 1990, http://cxc.harvard.edu/toolkit/colden.jsp); 
(2) $nhtot$ (using Willingale et al. 2013,
http://www.swift.ac.uk/analysis/nhtot/index.php). 
Willingale et al. (2013) calculate the molecular hydrogen column density, N(H2), in the Milky Way Galaxy
using a function
that depends on the product of the atomic hydrogen column density, N(HI), 
and dust extinction, E(B-V), with the aid of the 21 cm radio emission maps and the
\swi\ GRB data. Using these software, we found \nh\ in a range,  
0.6-0.7$\times$10$^{21}$ cm$^{-2}$,
0.8-1.2$\times$10$^{21}$ cm$^{-2}$, and 3.1-4.3$\times$10$^{21}$ cm$^{-2}$ for MV Lyr, BZ Cam and 
V592 Cas, respectively. Our \nh\ values are consistent with interstellar \nh\
within 90\% confidence level errors. The fits on Table 3 are performed over a range of binning
between 30-60 counts per bin, checking the soft X-ray range and the 
tabulated values and errors of \nh\  down to 0.4 keV and the hard X-ray band out to 7.5 keV.

We note that there was a preprint of a conference proceeding put in the archive 
(astro-ph) by Zemko \& Orio (2013)
including some preliminary XRT analysis of two of the sources BZ Cam and MV Lyr in our study 
which was after the submission of this paper to the $Astrophysical\  Journal$. These authors
use $Raymond$-$Smith$ and APEC models that are similar to the MEKAL models used in this study 
but do not use the expected multi-temperature distribution plasma models
(e.g. CEVMKL, MKCFLOW) 
as discussed in this section. 

Finally, we note that though some acceptable fits may be achieved with
higher \nh\  (e.g. using partial covering absorbers ect.) than interstellar \nh\ 
values using the \swi\ data, 
such models with higher \nh\  
do not reproduce the detected $ROSAT$ count rates for these sources
by factors of 3-100 times. Thus, they are not consistent with a global model for our targets 
in the 0.1-10 keV (or 0.1-50 keV) range.

\subsection{On the existence of blackbody emission components}

We do not find blackbody emission components consistent with our XRT spectra in our modeling. For completeness,
we exploit the data to calculate 2$\sigma$\ upper limits on temperatures and fluxes for
blackbody emission consistent with BL emission 
in which we have reduced the binning in the data and included a softer energy range down to 0.4 keV.
We calculate (1) in a range of kT$_{BB}$=(20-50) eV, an upper limit on the soft X-ray luminosity (blackbody) 
6.6$\times$10$^{32}$ \ergsec\ for MV Lyr in the 0.1-10.0 keV band; (2) in a range of kT$_{BB}$=(25-50) 
eV, a soft X-ray 
luminosity upper limit is 5.2$\times$10$^{32}$ \ergsec\ in the 0.1-10.0 keV band consistent with V592 Cas;
(3) For  kT$_{BB}$=30 eV upper limit, we find 
a soft X-ray luminosity upper limit of  2.3$\times$10$^{33}$ \ergsec\ in the 0.1-10.0 keV band
for BZ Cam. We can not calculate
upper limits on the soft X-ray flux/luminosity for effective temperatures lower than the given values/ranges iin the above calculation.
For larger effective temperatures than the given values/ranges, the flux/luminosity upper limits 
are at lower values (i.e., $\le$ 1$\times$10$^{32}$ \ergsec\ ) than these upper limits. 
We caution that these upper limits
are not better than $ROSAT$ results in the literature
due to the energy band width of \swi\ and its soft sensitivity. We stress that
the comprehensive $ROSAT$ analysis of the 11 nonmagnetic NL systems, 
with its better soft X-ray sensitivity in the 0.1-2.4 keV band,
demonstrates that the hardness ratios of these sources are not compatible with 
blackbody emission
(cf. Fig 2 from van Teeseling et al. 1996) and they are consistent with optically thin plasma emission.
A joint $ROSAT$ and $EUVE$ data analysis of IX Vel (another NL) shows no blackbody emission 
(van Teeseling et al. 1995).

Therefore, we aquired the archival $ROSAT$ data of MV Lyr (PSPCB detector), BZ Cam (PSPCB detector), and 
V592 Cas (PSPCC detector),
and analyzed the data using the model of $tbabs$$\times$CEVMKL and $tbabs$$\times$(BBODY+CEVMKL) 
utilizing the response files
{\it pspcc-gain1-256.rsp} and {\it pspcb-gain2-256.rsp}. 
We used observations obtained in the high states. The main goal of this analysis
was to derive possible 2$\sigma$\ temperature upper limits
for any soft blackbody emission consistent with BLs.
Our fits using the CEVMKL model yielded acceptable \rchisq\ values (\rchisq=1.0-1.2)
with parameters close to the best-fit spectral parameters and in the error ranges given in Table 3. 
We do not present these since the derived results are similar with the ones in Table 3. 
The $\alpha$ parameter for the
power law index of the temperature distribution in CEVMKL model was fixed at the best fit values in 
Table 3 for  BZ Cam and V592 Cas.
We note that we have not included the power law spectral component in the fits since the $ROSAT$ 
fits were conducted in 0.1-2.2 keV
band, and this model will not be detected along with the thermal plasma emission component, 
in the narrow energy  band. 
In order to calculate the 2$\sigma$ blackbody temperature upper limits, we added a second component of 
BBODY (in XSPEC) in the modeling. The fits
yielded the following results: (1) MV Lyr, kT$_{BB}<$ 6.6 eV; (2) BZ Cam, kT$_{BB}<$ 5.4 eV; and (3) V592 Cas, 
kT$_{BB}<$ 7.1 eV.
These 2$\sigma$ blackbody temperature upper limits are consistent with the detected or expected WD temperatures in these NLs and
not with any optically thick boundary layer emission. We note that if single MEKAL models are used for the 
fitting, the derived temperature upper limits are very similar. 

\section{Discussion} 

NLs have not been thoroughly investigated using the recent X-ray telescopes with higher sensitivity,
wider energy band and better energy resolution in comparison with older mission studies of $ROSAT$ and
$EINSTEIN$. We observed a selected group of 
non-magnetic NLs studied only with $ROSAT$ in a narrow energy band to further investigate 
their spectral and temporal characteristics using \swi. 
X-ray observations of BLs in NL systems are important to study characteristics under the effect of 
high accretion rates,
to investigate the geometry of the inner disk, to calculate the BL temperature, BL luminosity, 
and mass accretion rate. With simultaneous X-ray and UV observations, 
one can work on a self-consistent model of the BL and the accretion disk.
 
We have presented the \swi\ XRT spectral and temporal analysis of three NL systems,
MV Lyr, BZ Cam and V592 Cas.
The temporal analysis indicates that all systems show X-ray variations over the binary orbit
as observed in  the mean light curves.
We suggest this may be related to scattering over the accretion disk. This may result
from a modulated disk wind 
or an elevated disk rim in the accretion impact zone.
However, the latter requires high inclination that is not consistent with these sources. Thus,
scattering from disk winds that are episodic and modulated with the orbital period,
is likely to create the X-ray variations over the orbit in accordance with the wind activity. 
However, it is important to
note that the variations seem to follow the wind outflows as they maximize with the orbital modulation 
of the H$\alpha$ line 
which also shows P Cygni profiles (see V592 Cas). 
This indicates that the X-ray emitting regions may be at the
base of these strong wind outflows. There is a need for better/longer X-ray time series data with 
simultaneous photometric light curves in the optical and UV for direct comparisons. 
 
The \swi\ XRT spectral analysis of the three sources reveals hard X-ray emission with
high X-ray temperatures in the 0.2-10.0 keV range.
We found an X-ray temperature kT$_{max}$
 $>$21 keV for MV Lyr, 33.3$^{+16.0}_{-14.0}$ keV for BZ Cam and 35.5$^{+19.7}_{-10.9}$ 
keV for V592 Cas. 
Amoung the three sources MV Lyr has the hardest X-ray spectrum, thus the maximum limit of the
temperature was not constrained in the \swi\ band (see Table 3 for a best fit value).
Corresponding X-ray luminosities of the thermal plasma emission 
are 2$\times 10^{31}$ \ergsec, 
2.2$\times 10^{31}$ \ergsec\ and 5.3$\times 10^{31}$ \ergsec, respectively, calculated from the CEVMKL
model (i.e., fitted model) parameters. The total X-ray luminosities (thermal+power law)
are 1.7$\times 10^{32}$ \ergsec\ for MV Lyr, 4.6$\times 10^{32}$ \ergsec\
for BZ Cam, and 5.3$\times 10^{31}$ \ergsec\ for V592 Cas in the 0.2-10.0 keV energy band. 
We calculated that in the 0.1-50 keV band (where bulk of the X-ray emission is) the
X-ray luminosities of the thermal CEVMKL components are
1.9$\times 10^{32}$ \ergsec, 5.0$\times 10^{32}$ \ergsec, and 8.5$\times 10^{31}$ \ergsec, for
MV Lyr, BZ Cam, and V592 Cas, respectively.  
The fits for MV Lyr and BZ Cam involves a second spectral component. 
In order to double check the plausible under estimation of the thermal plasma emission luminosity,
we assumed a single CEVMKL component fit for these two sources. We fitted the $tbabs$$\times$CEVMKL
model fixing the maximum plasma emission temperatures from the fits in Table 3, which resulted acceptable
\rchisq. We re-calculated the 
thermal plasma emission luminosity in the 0.1-50 keV band for these sources 
and found 3.2$\times 10^{32}$ \ergsec, 
and 7.0$\times 10^{32}$ \ergsec\ for MV Lyr and BZ Cam, respectively. 
Thus, we are not underestimating the thermal plasma luminosities from the optically thin boundary layers.
The X-ray flux and luminosity of the power law components,
may change considerably in the 0.1-50 keV band 
due to the power law index and the cut-off energy from the thermal electron 
distribution. Thus, it needs to be determined from fits in a larger
energy band width. We note that power law components are not part of the transfered 
accretion power of the systems but are related to additional scatterings.
 
As summarized in the introduction, the X-ray emission from about 11 NLs have been studied with 
$ROSAT$  and found to show
optically thin emission with a single MEKAL model at temperatures $<$6 keV (best fit values) and
luminosities $<$10$^{32}$ \ergsec\ . A few NLs were studied with $ASCA$ and modeled with double MEKAL models
at different temperatures (e.g., TT Ari and KT Aur, Mauche \& Mukai 2002), and one
with \xmm\ using multiple-temperature plasma and MEKAL models (Pratt et al. 2004).
Note that in all these observations the X-ray luminosities are $\le$ afew $\times$10$^{32}$ \ergsec.
Greiner (1998) found blackbody model of emission more consistent with $ROSAT$ spectra
with (0.2-0.5) keV temperatures and hydrogen column densities of interstellar values 
however, with very small emitting regions. Other than these blackbody temperatures, $ROSAT$ (0.1-2.4 keV)
does not reveal any blackbody component in NL spectra given the suitable soft X-ray energy band and 
its sensitivity. In addition, the $ASCA$ and \xmm\ data
do not reveal these hot blackbodies or any blackbody emission. 
Most of these results are in accordance with our findings
and we find hard X-ray spectra with multi-temperature 
optically thin plasma emission  
in the \swi\ energy band without a blackbody emission, component consistent with $ROSAT$ 2$\sigma$
upper limits for blackbody emission with kT $<$ 7 eV.

We have calculated the accretion rates in the hard X-ray emitting BLs of our  
systems (L$_{BL}$=GM$\dot{\rm M}$/2R$_{WD}$).
The accretion rate in the boundary layer is 6.7$\times 10^{-11}$ \msol\ yr$^{-1}$ for MV Lyr, 
1.4$\times 10^{-10}$ \msol\ yr$^{-1}$ for BZ Cam and 1.9$\times 10^{-11}$ \msol\ yr$^{-1}$ for V592 Cas.
For these calculations, values of M$_{WD}$ and R$_{WD}$ are assumed from Table 1 (for BZ Cam
a similar WD mass and radius is taken as in MV Lyr) and thermal plasma emission
luminosities in the 0.1-50 keV band have been used.
It is of importance to compare our observational results with 
theoretical calculations of standard steady-state
disk models, e.g. of Narayan \& Popham (1993), and Popham \& Narayan (1995). 
Taking the accretion rates calculated for these NLs in the optical and 
UV bands 1.3$\times 10^{-8}$ \msol\ - 3.9$\times 10^{-9}$ \msol\ (see Table 1), the standard disk models 
mostly predict optically thick BLs showing a blackbody
emission spectrum with temperatures of 13-33 eV and L$_{soft}$$\ge$ 1$\times 10^{34}$ for an 
0.8-1.0 \msol\ WD, (WD rotation as fast as $\Omega_{*}$=0.5$\Omega_K(R_{*})$ is taken into 
account in L$_{soft}$, see Popham \& Narayan (1995)).
We underline that this blackbody temperature range is definitely larger than the 2$\sigma$
upper limits we derive from the $ROSAT$ data and the luminosity range seems larger than the
luminosity range (0.1-10.0 keV) we calculated from the \swi\ data (see section 6.1).  
The standard steady-state constant $\dot{\rm M}$\ disk models 
predict (for optically thin emission) X-ray temperatures around 9-10 keV for a 1 \msol\ WD
at about  3.2$\times 10^{-10}$ \msol (see Narayan \& Popham 1993). 

Comparing the accretion disk luminosity from Table 1 and the X-ray luminosity 
of the thermal plasma emission from these three NLs, 
(1.8$\times 10^{32}$ \ergsec, 5.0$\times 10^{32}$ \ergsec and 8.5$\times 10^{31}$ \ergsec, for
MV Lyr, BZ Cam, and V592 Cas, respectively) we find that the ratio 
L$_{x}$/L$_{disk}$ $\sim$ 0.01-0.001\ .
The accretion rates and luminosities calculated using the X-ray data of these NLs resemble dwarf novae 
in quiescence. However, the optical and UV rates and luminosities for NL disks in
the same brightness states resemble those of dwarf novae in outburst. 
For nonmagnetic systems, the earlier studies of the ratio of the X-ray flux to optical and/or UV flux, 
$F_x/F_{opt}$, decreased along the sequence SU UMa stars ($F_x/F_{opt}$$\sim$0.1)- U Gem stars- Z Cam
stars ($F_x/F_{opt}$$\sim$0.01) - UX UMa stars and high state VY Scl stars ($F_x/F_{opt}$$\le$0.001) 
using the $EINSTEIN$ and $ROSAT$ results 
(see Kuulkers et al. 2006, van Teeseling et al. 1996, Patterson \& Raymond 1985). This ratio is found to
decrease with increasing P$_{orb}$ and increasing $\dot{\rm M}$\ where a high $\dot{\rm M}$\ 
causes the disk to emit more UV flux, but not as much X-ray flux (see review by Kuulkers et al. 2006).

The virial temperature in the inner parts of the accretion disk 
(kT$_{virial}$=$\mu$m$_p$GM$_{WD}$/3R$_{WD}$ where $\mu$$\sim$ 0.6 and m$_p$ is the 
proton mass) limits the maximum plasma emission temperatures 
in nonmagnetic CVs (see also Pandel et al. 2005 for dwarf novae in quiescence).
All the kinetic energy from the Keplerian motion of the accreting gas
is converted into heat when virial temperatures are aquired by the flow.
This gives the maximum energy per particle that can be dissipated and 
the temperature of standard 1-D boundary layers cannot be around the virial temperatures 
or material can not be confined to the disk. Given the WD properties on Table 2, if one assumes similar
properties for the three NLs in question, we calculate T$_{virial}$=27 keV for these NLs. 
Energy budget of a radiative isobaric cooling flow indicates that T$_{max}$
is smaller than T$_{virial}$. The cooling flow releases an energy of (5/2)kT$_{max}$
per particle including kinetic and compressional components. The total thermal/kinetic
energy at the inner edge of the disk available per particle is (3/2)kT$_{virial}$.
This inturn yields another constraint that T$_{max}$/T$_{virial}$$<$ 3/5.
Therefore, given T$_{virial}$=27 keV, then T$_{max}$$\le$ 16 keV. 
All of the best fit X-ray temperatures are above the calculated virial temperature
and T$_{max}$ constraint. We also calculated the 2$\sigma$ lower
limits of the X-ray temperatures to compare with our T$_{virial}$ and T$_{max}$ 
estimations and find that for BZ Cam the limit is $>$ 17 keV and for MV Lyr and BZ Cam, it is $>$ 21 keV. 
As a result,  the boundary layers in these NLs are too hot, thus
may not be confined to the disk and will expand/evaporate 
forming  {\it \bf ADAF-like flows/X-ray coronal regions}
where the accreting material will be advected onto the WD. 
We stress that all the multi-temperature plasma model fits
show that the $\alpha$ parameter in these fits 
differ from 1.0 and that the cooling flows are non-standard as well revealing a different type of flow.

ADAFs (Advection-dominated flows) correspond to a condition where the gas is radiatively inefficient and the
accretion flow is underluminous (see Narayan \& Mc Clintock 2008, Lasota 2008, Done et al. 2007).
Advection-dominated accretion may be described in two different regimes. 
The first is when the accreting material has a very low density 
and a
long cooling time (also referred as RIAF-radiatively inefficient accretion flow) 
with t$_{cool}$ $>$ t$_{acc}$
(t$_{cool}$  is the cooling time and t$_{acc}$ is the accretion time). This causes the accretion flow 
temperatures to be virialized in the ADAF region. 
Note that the standard ADAFs are two-temperature flows as studied in LMXBs
where electrons and ions are at differing temperatures. 
The electron temperatures increase with decreasing accretion rate.
For example, Atoll type relatively low accretion rate (compared to Z-sources) 
neutron star LMXBs (type-I bursters)
show 1-200 keV spectra consistent with thermal Comptonization  
with a plasma electron temperature around 25-30 keV
and some are found at even lower X-ray temperatures 
(see Barret et al. 2000 and Done et al. 2007). These systems
show an optically thin BL as the main region of energy release merged with an ADAF. 
These sources have
similar accretion rates to NLs if the ratio to $\dot{\rm M}_{\rm Edd}$ is considered in relative terms.
Note that neutron stars in Atoll sources should be accreting substantial material as revealed in their X-ray bursts.
For LMXBs hosting black holes the ADAFs have detected temperatures about 100 keV. For CVs
virial temperatures in the disk are around 10-45 keV where the 
WDs are primaries as opposed to neutron stars or black holes (assuming 0.4 \msol\ - 1.1 \msol WDs). 

The second regime is such that
the particles in the gas can cool effectively, 
but the scattering optical depth of the accreting material is large enough
that the radiation can not escape from  the system 
(see also "slim disk" model Abramowicz et al. 1988). The defining
condition is then, t$_{diff}$ $>$ t$_{acc}$ (t$_{diff}$ is the diffusion time for photons). 
This regime requires high accretion rates of the order of 0.1$\dot{\rm M}_{\rm Edd}$. 
We stress that the mass accretion rates derived from the optical and UV observations
for the three NLs in this 
paper are below this critical limit for a slim disk approach.

In an RIAF-ADAF region (first regime) energy liberated
by viscous dissipation remains in the gas and the energy is advected onto the 
compact star given some ratio of advection energy and viscous dissipation. 
Thus, the pressure and hence the sound speed are large. The accretion flow becomes
geometrically thick, with high pressure support in the radial direction which causes 
the angular velocity to stay at
sub-Keplerian values, and  the radial velocity of the gas becomes 
relatively large with $\alpha$$\sim$0.1-0.3. 
This leads to a short accretion time t$_{acc}$$\sim$R/v$\sim$t$_{ff}$/$\alpha$ (t$_{ff}$: free fall timescale;see Narayan \& Mc Clintock 2008). 
Finally, the gas with the large velocity and scale height
will have low density, since the cooling time is long , and the medium will be optically thin.

It is widely accepted that there is strong connection/association between ADAFs and outflows as in winds
and jets/collimated outflows (see Narayan \& Yi 1995, Blandford \& Begelman 1999). 
This is largely because the ADAFs have positive Bernoulli parameter
defined as the sum of the kinetic energy, potential energy and enthalpy, thus the gas is not well bound 
to the central star. We point out that alot of the NL systems have strong wind outflows 
and particularly,  two of the NLs studied in this paper, BZ Cam
and V592 Cas, have strong bipolar collimated wind outflows modulated with the orbital period (e.g.,  
rotates obliquely to the observer`s line of sight), as detected in the optical 
and UV bands with high velocities 3000-5000 km s$^{-1}$ (see Introduction). 
Given these characteristics, we suggest that the ADAF-like optically thin BLs may be the origin of 
these wind outflows 
from these systems. Mass loss rates of winds in NLs are about or less than 1\% of the accretion rate,
with acceleration length scale of about (20-100)R$_{WD}$ which are strongly affected by 
rotation (Kafka \& Honeycutt 2004, Long \& Knigge 2002). 
For V592 Cas, as noted earlier in the Introduction, optical brightness variations
are not correlated with the strength of the wind outflows, and the outflow shows orbital 
variations and no variations
relating to a disk tilt, disk precession or any superhump period of the system. Moreover, in the UV 
emitting inner disk, \citet{fro12} finds that the continuum does not vary in a good fraction of the 
target NLs, revealing that changes in the wind 
structure are not tied to the accretion disk. These are a few hints that
the origin of the wind outflow may be in the ADAF-like BL, not the disk itself. 
We note that BZ Cam is a CV embedded in a nebula not related to a nova explosion.
It has been suggested that interactions of the BZ Cam wind with the
interstellar medium produces the observed nebular bow shock
(Hollis et al. 1992), but alternative origins for the nebula have
also been proposed (e.g., Griffith et al. 1995; Greiner et al. 2001). 

Narayan \& Popham (1993) show that the optically thin BLs of accreting WDs
in CVs can be radially extended and that they advect part of the
viscously dissipated energy as a result of their inability to cool, therefore 
indicating that optically thin BLs act as ADAF-like accretion flows. Partial radial pressure support 
by the hot gas also results in sub-Keplerian rotation profiles 
in these solutions. Advection can play an important role for the energy budget and the emission 
spectrum of the accretion flow in the BL. In addition, Popham \& Narayan (1995) illustrate that
the BL can stay optically thin even at high accretion rates for optical depth $\tau$ $<$ 1 together with
$\alpha$ $>$ 0.1\ . However, such models are not well investigated.
For simplicity an ADAF around a WD can be
described by  truncating the ADAF solutions (as opposed to BHs) 
and the accretion energy is advected onto the WD
heating it up. Accretion via a standard 
disk boundary layer is expected to spin-up a WD 
except when rotating near breakup (Popham \& Narayan 1991), however a WD is spun down when 
accreting via ADAF-like hot flows (Medvedev \& Menou 2002). Sion (1999) emphasizes that rapidly rotating
WDs are rare in nonmagnetic CVs.
For some preliminary modeling of ADAFs or hot settling flows (for CVs) 
see Medvedev \& Menou (2002) and related references therein. 
The ADAF models in CVs should utilize
reemission of the energy advected by the flow, modeled as a single-temperature blackbody of temperature:
T$_{eff}=(L_{adv} / 4f\pi\sigma R_{wd}^2)^{1/4}$
where f is the fraction of the stellar surface that is emitting. In addition, we 
note that Godon \& Sion (2003)
have calculated that WDs in DN are not generally heated to more than 15\% above their original temperature
via standard steady-state disk accretion, even through multiple DN outbursts. Hence, other mechanisms are 
necessary (e.g. compressional heating, heating via advective accretion flows, irradiation via
luminous optically thick boundary layer).
Characteristics of ADAF-like flows may be different in CVs,
since WDs are less compact than stellar-mass BHs, the hot flows will have less extend
around a WD. Menou (2000) suggests that
the gas in the hot flow is one-temperature (1-T implying plasma in collisional ionization equilibrium--CIE), 
because of the efficient Coulomb interactions at 
lower temperatures
compared to BHs and that the energy advection is not dependent on the 
preferential heating of the
ions by the viscous dissipation and lack of energy exchange between the ions and the electrons necessary for
the two-temperature ADAFs in BH binaries (2-T implying a type of non-equilibrium ionization plasma--NEI). 
We note here that
since there are no detailed theoretical calculations for ADAF flows in CVs, existence of 2-T ADAFs
for CVs can not be completely ruled out which may yield a rather featureless X-ray spectra.
The energy advected by the flow is lost through the event horizon
in the BHs whereas in the CVs, it is expected to be re-radiated from the surface of the WD. 
The resulting EUV emission (and/or hard X-ray emission from the BL) 
could explain the strength of HeII $\lambda$4686 emission line as due to
disk irradiation. 

One can compare the WD temperatures/heating 
in magnetic CVs (MCVs) and NL systems. This is only to calculate an approximate radiation efficiency
for the BLs in NLs  since MCVs particularly Polars,
have different accretion geometry (they do not have accretion disk) 
and thus, does not accrete through  advective hot flows.
MCVs are known to have radiative shocks in accretion columns over the magnetic polar caps
heating the WDs through basically radiative accetion flows. 
The Polar subtypes have been found to host cooler WDs compared with nonmagnetic CVs at the same
orbital period (Sion 1999, Araujo-Betancor et al. 2005, Townsley \& Gansicke 2009).
The accreted material in these systems (possibly valid for IPs) 
are constained by the magnetic field to the polar regions down to a limited 
pressure depth. Strong lateral pressure gradients forces this material 
to spread over the surface. As a result, heat dissipated by compression up to a critical pressure 
is constrained to the polar regions, 
and compression at further higher pressures is spread over the whole surface heating the WD
(Townsley \& Gansicke 2009). 
The average WD temperature is about 50,000 K and about 16,000 K in NLs and MCVs (Polars), respectively
above the period gap between 3-5 hrs (see Townsley \& Gansicke 2009, Mizusawa et al. 2010). 
The relative luminosity difference
of the WDs as a result of these temperatures is L$_{adv}$/L$_{rad}$ = (50000/16000)$^4$ 
(assuming similar R$_{WD}$).
This ratio is about 96 which means that about 96 times more accretion power may be transferred to the WD
in NLs. This will approximately result in  about a factor of 96 times reduction in the radiation 
power from the BL since this power is advected onto the WD yielding a radiation inefficiency in the BL 
of about 0.01; we may denote this as $\epsilon_{adv}$.  We note that if one assumes average temperature
from the total samples of Polars and NLs, L$_{adv}$/L$_{rad}$ = (45000/13500)$^4$ =123, yielding
an approximate $\epsilon_{adv}$ $\sim$ 0.008\ (same average R$_{WD}$ assumed).

As we have discussed in the previous paragraphs it has been proposed that ADAFs can have 
thermally-driven winds (Narayan \& Yi 1995, Blandford \& Begelman 1999) possibly
modified/controlled by magnetic fields
that would carry away some of the accretion power in the BL which is, then, not radiated by the BL.
This does not necessarily imply that all NL winds are to originate from the BL. 
Therefore, the power/energy necessary to derive the wind outflows from BLs will also 
yield inefficiency in the 
radiation from the BL in the NLs ; we may denote this as $\epsilon_{wind}$. 
The wind  luminosity, which represents the kinetic energy loss of the wind, 
is described as L$_{wind}$=$(\dot{M}_{wind}v_\infty^2)/2$. Not all the power used to derive the
wind outflows appears as the kinetic power of the outflow and there is a certain efficiency factor
associated with this conversion that changes from typically 10$^{-1}$-10$^{-4}$ 
for stellar winds (Lamers \& Cassinelli 1999) or 10$^{-3}$-10$^{-4}$ for AGNs (Yuan \& Narayan 2014).
Assuming 1\% of 
the  power used to drive the wind outflows appears as the kinetic power of the outflow 
defined as the wind luminosity,
L$_{outflow-BL}$=L$_{wind}$/0.01 . Using a wind speed of 3000 km s$^{-1}$ and a mass loss rate of 
1$\times 10^{-10}$ \msol\ yr$^{-1}$ (Kafka et al. 2009), 
L$_{wind}$ is 3$\times 10^{32}$ \ergsec\ which will reduce 
the BL radiation by about 3$\times 10^{34}$ \ergsec\ in V592 Cas. The same calculation using
5000 km s$^{-1}$ and 1$\times 10^{-11}$ \msol\ yr$^{-1}$ (Honeycutt et al. 2013) yields a reduction
in the  BL radiation by about 8$\times 10^{33}$ \ergsec\ in BZ Cam. The wind in MV Lyr is not as fast
(see Linnell et al 2005) and does not yield strong reduction ($\le$2$\times 10^{31}$ \ergsec). 
Narayan \& Yi (1995) predict that diminished winds/outflows may possibly occur 
when a hot (disk) corona forms (as viewed in the X-rays) with less effective advection-dominated flows.
Overall, the efficiency of the BL radiation will be
reduced by $\epsilon_{total}$=$\epsilon_{adv}$$\times$$\epsilon_{wind}$.

It is possible that the soft X-ray to EUV emission from the region of the BL and/or the inner disk
of a steady-state disk can be screened by the existing ADAF-like/X-ray coronal region and the
soft radiation from the disk can be Compton-upscattered. 
Another possibility is that hard X-ray photons from the BL can Compton/scatter
from the existing (strong) winds in these systems. Some of these winds are axisymmetric/bipolar in
nature which may lead to non-isotropic scattering.
Thus, the power law components derived from our fits are crucial and signals the existence
of Comptonized or Compton up-scattered radiation in
these systems as described above. 
It may be possible that both type of thermal Comptonization or scattering is occurring in these systems.
We find power law luminosities of (0.2-2.4)$\times$10$^{32}$ \ergsec 
(consistent with only some
percent of the total X-ray luminosities of these NLs). 
We caution that the power law components
also do not account for the disk luminosity of these systems (they are about  
1-0.1\% of the disk luminosity).

There has been no definitive evidence  
that NLs are magnetic CVs; a few were suggested as intermediate polars
(IP) (see \cite{fro12}), but have never been confirmed in X-rays. The UV and optical
spectra of NLs are different from those of IPs. In particular,  episodic bipolar  
strong wind production is
not seen in the magnetic systems. The WDs in NL systems are very hot,  
2-4 times hotter (see Townsley \& Gansicke 2009, Mizusawa et al. 2010 and see also Table 1) in general 
compared with magnetic CVs revealing differences in the accretion physics,
heating and geometry. 
The X-ray emission in IPs show virialized hot plasma from stand-off radiative shocks in small
regions (accretion column) around the magnetic poles
with temperatures 10-95 keV at 10$^{31-34}$ \ergsec 
(Balman 2012 and references therein). On the other hand, the most important  signature of IPs 
is the spin period of the WDs that manifests
itself in the X-ray light curves regardless of the inclination of the systems. 
None of our 
systems (and most other NLs) have  detected spin periods in the X-rays and 
we did not detect any 
(see Mukai 2011 for a catalog of IPs \footnote[1]{ 
http://asd.gsfc.nasa.gov/Koji.Mukai/iphome/catalog/alpha.html}).
IP X-ray spectra show complex absorption phenomenon with partial covering absorption or
warm absorbers showing energy dependent modulations where we only found interstellar 
absorption and no energy dependence of orbital variations atypical of IPs. 
Moreover, IPs with high 
accretion rates, as detected from the optical and UV bands of these NLs 
may be expected to show a component with a 40-120 eV blackbody temperature
resulting from reprocessing of the X-rays, for which we find no evidence.
About 30\% of IPs indicate this soft component (Bernardini et al. 2012). In addition,
IPs do not show major high and low states as in two of our NLs.  
Thus, there is no consistency with a magnetic CV picture for our targets.

\section{Conclusions}

Observations of CV disk systems at low mass accretion rate
(namely dwarf nova CVs in quiescence) 
have yielded the temperature and luminosity of the BLs from X-ray observations while the
UV observations helped to determine the temperature and luminosity
of the accreting white dwarfs, as they are the dominant
component at low $\dot{M}$. The quiescent X-ray spectra are well characterised with a multi-temperature 
isobaric cooling flow model of plasma emission at kT$_{max}$=6-55 keV with accretion rates of
10$^{-12}$-10$^{-10}$ M$_{\odot}$ yr$^{-1}$ and the detected Doppler broadening in 
lines during quiescence is $<$750 km s$^{-1}$ at mostly sub-Keplerian velocities
with electron densities $>$10$^{12}$ cm$^{-3}$. To further our understanding of the boundary layers
in CVs and their spectral and temporal characterictics in the X-rays, we used a group of
non-magnetic NLs (VY Scl and UX UMa sub-type systems)
where NLs are  mostly found in a state of high mass accretion rate 
with $\ge$1$\times$10$^{-9}$ \msol\
yr$^{-1}$ and some showing occasional low states (VY Scls). 
X-ray observations of BLs in NL systems are important to
derive parameters such as the BL temperature, BL luminosity, mass accretion rate and study
accretion geometry. Using  both X-ray and UV observations
one can then build a self-consistent model of the BL and the accretion disk.

In this work, we presented \swi\ XRT observations of three non-magnetic NLs, 
MV Lyr (VY Scl type), BZ Cam (VY Scl type),
and V592 Cas (UX UMa type), obtained in their high states. 
We find that these sources have X-ray spectra
consistent with multi-temperature hot plasma emission that has a power law dependence in emission measure 
(i.e. temperature) distribution in
a range kT$_{max}$=(21-50) keV. We calculate that the X-ray emitting plasma is virialized.
The power-law index of the temperature distribution indicates that these plasmas depart 
from the predictions of isobaric cooling flow type models as opposed to low $\dot{M}$ systems 
like quiescent DN. 
We detect a second component in the X-ray spectra that is well modeled by a
power law emission in the VY Scl type sources BZ Cam and MV Lyr. V592 Cas may be fitted
with an additional power law model, but the fit is not significantly different than
a fit without it. 
We do not find any periodicity in the \swi\ XRT data for the three sources, but we find non-sinusoidal
variation of X-ray emission over the orbital cycle of the systems without energy depence 
in the 0.2-10.0 keV band. Particularly, the mean light curve of V592 Cas 
indicates two seperated peaks that may be associated with a bipolar outflow structure.
The ratio of the unabsorbed X-ray flux/luminosity of the three systems 
and the disk luminosities as calculated from UV and optical wavelengths is $\sim$0.01-0.001
where BZ Cam and MV Lyr are in the upper end and V592 Cas is in the lower end of this range indicating
inefficiency of X-ray/BL radiation (note that we quote a lower limit for the disk luminosity of BZ Cam
in Table 3).
We do not find a blackbody emission component in the soft X-rays using our \swi\ data and also the $ROSAT$
data with a 2$\sigma$ upper limit of kT$_{BB}$ $<$ 7 eV for the three systems, consistent with the
WD temperatures but not with the optically thick boundary layer emission.     

As a result, we suggest that the BLs in NL systems may be optically thin hard 
X-ray emitting  regions merged with ADAF-like flows and/or constitute X-ray corona 
regions on the inner disk close to the WD. 
We estimate that the high WD temperatures in the three NLs and others (NLs) may explain the
efficiency reduction ${\epsilon}_{adv}$ in the optically thin BLs with a factor $\sim$ 0.01\ . 
ADAF-like accretion flows
in the BLs may help to explain the very fast collimated outflows from the NL systems 
because ADAFs have positive Bernoulli parameter
(the sum of the kinetic energy, potential energy and enthalpy). Thus,
the power lost to drive a wind from the ADAF-like BLs may result in even larger
losses from the accretion power in these X-ray emitting BL regions.
In addition, we note that analysis in the optical and the 
UV wavelengths (in the IR for some)  
indicate departure from the standard disk model for MV Lyr (Linnell et al. 2005),
BZ Cam (Godon et al. 2014, in preparation), and V592 Cas (Hoard et al. 2009). 
We caution that our findings in the X-rays  may not be optimally 
analogous to the case of LMXBs
thus, ADAF-like flows (1-T in CIE or plausibly 2-T in NEI ) and 
formation of X-ray coronal regions in the BLs and/or the disk
needs to be modeled for CVs in detail. We can not confirm at this stage that all NL systems 
have the type of BL/inner disk structure described for these three sources in this paper.

\section*{acknowledgment} 
The authors thank J. P. Lasota, J. Greiner and M. Revnivtsev for careful 
comments on the manuscript.
PG is thankful to William Patrick Blair at the 
Henry Augustus Rowland Department 
of Physics and Astronomy at the Johns Hopkins University (Baltimore, MD),  
for his kind hospitality. 
This work was supported by the National Aeronautics and Space Administration
(NASA) Under grant number NNX13AJ70G issued through the Astrophysics
Division Office (SWIFT Cycle 8 Guest Investigator Program) to 
Villanova University.

\newpage   

\begin{deluxetable}{lllll}
\tablewidth{0pt}
\tablecaption{System Parameters}
\tablehead{
Parameter  & units      & MV Lyr   & BZ Cam   & V592 Cas               
}
\startdata
$M_{wd}$ & $M_{\odot}$             & 0.73-0.8$^{~(1,2)}$      &    & 0.75$^{~(11)}$   \\ 
$R_{wd}$ & $km$                    & 7,440$^{~(3)}$           &    & 7,378$^{~(11)}$   \\ 
$M_{2nd}$ & $M_{\odot}$            & 0.3$^{~(2)}$             &    & 0.21$^{~(11)}$   \\ 
$i$ & $deg$                        & 10$\pm3^{~(4,5,6)}$      & 12-40$^{~(13,14)}$   & 28$\pm10^{~(10)}$   \\ 
$P$ & $hr$                        & 3.19$^{~(5)}$            & 3.69$^{~(9,17,18)}$ & 2.76$^{~(8)}$   \\
$d$ & $pc$                         & $505\pm 30^{~(1,2)}$     & 830$\pm$160$^{~(13)}$    & 330$^{~(10)}$-360$^{~(11)}$   \\
$T_{wd}$ & $K$                     & $44,000\pm3000^{~(1,2)}$ &           & $45,000^{~(11)}$ \\
$\dot{M}_{high}$ & $M_{\odot}$/yr$$ & $3 \times 10^{-9~(1,6)}$ &  $\ge$$3 \times 10^{-9~(18)}$  & $\sim 1.3 \times 10^{-8~(11)}$ \\
V               & min-max          & 17.7-12.1                & 14.3-12.5 & 12.9-12.5   \\ 
L$_{disk}$  & \ergsec         & $\sim 2.7 \times 10^{34}$  & $\ge$$3 \times 10^{34}$  & $\sim 1.2 \times 10^{35}$ \\
$\Omega_*$ & $\Omega_K$            & $\sim$0.28$^{~(1)}$      &           &    \\
\enddata  \\ 
\flushleft{
{\bf References:} 
(1) \citet{god12}; 
(2) \citet{hoa04}; 
(3) \citet{woo95};  
(4) \citet{sch81};    
(5) \citet{ski95};     
(6) \citet{lin05}; 
(7) \citet{bru94}; 
(8) \citet{tay98};  
(9) \citet{pat96};  
(10) \citet{hub98};  
(11) \citet{hoa09}; 
(12) \citet{car89}; 
(13) \citet{rin98}; 
(14) \citet{lad91}; 
(16) \citet{pri00}; 
(17) \citet{hon13}; 
(18) \citet{tho93}; 
(19) \citet{ballz09}.
} 
\end{deluxetable}

\clearpage

\begin{deluxetable}{cccccrcc}
\tablewidth{0pt}
\tablecaption{{\it SWIFT} Observation Log}
\tablehead{
System & Obs ID & Seg & Date (UT)  & Time (UT)& Exp. Time & UVOT & XRT \\ 
Name   &        &     & yyyy-mm-dd & hh:mm:ss & seconds   & mode & mode                 
}
\startdata
MV Lyr   & 91443 & 4 & 2012-06-09 & 09:19:01 &  6600.00  &  0x122f & PC/WT \\ 
MV Lyr   & 91443 & 2 & 2012-06-08 & 09:04:01 &  7640.00  &  0x122f & PC/WT \\ 
BZ Cam  &  91441 & 2 & 2012-12-21 & 10:05:59 &  1240.00  &  0x122f & PC/WT \\ 
BZ Cam  &  91441 & 2 & 2012-12-21 & 10:09:59 & 13860.00  &  0x122f & PC/WT \\ 
V592 Cas & 91442 & 4 & 2012-09-10 & 09:49:59 &  6865.00  &  0x122f & PC/WT \\ 
V592 Cas & 91442 & 2 & 2012-09-09 & 04:44:45 &  6385.00  &  0x122f & PC/WT  
\enddata
\end{deluxetable}

\clearpage

\begin{deluxetable}{lllll}
\tablewidth{0pt}
\tablecaption{Spectral Parameters of the  
Fits to the NL Spectra}
\tablehead{
Model  & Parameter & MV Lyr & BZ Cam & V592 Cas           
}
\startdata 
CEVMKL & $N_H$(10$^{22}$atoms cm$^{-2}$) & $0.13^{+0.12} _{-0.06}$ & $0.30^{+0.07}_{-0.07} $ & $0.3^{+0.2}_{-0.2} $  \\     
       & $\alpha$              & $1.6^{+2.7}_{-0.4} $ & $0.13 ^{+0.16}_{-0.06} $ & $0.6^{+0.7}_{-0.3} $  \\    
       & $T_{max}$(keV)        &  $>$21$^{\dagger}$   & $ 33.0 ^{+16.0}  _{-14.0} $    & $ 35.5^{+19.7}_{-10.9} $ \\   
 & $K_{CEVMKL}$ & $9.2^{+7.0}_{-4.8} \times 10^{-4}$ & $6.7 ^{+2.8}_{-1.0} \times 10^{-4}$ & $2.3^{+1.1}_{-1.0}\times 10^{-3}$  \\ 
\hline
Power law & PhoIndex$_{powerlaw}$ & $0.82 ^{+0.07}_{-0.07}$           & $0.40 ^{+0.1}_{-0.3}$           &  $<$ 1.0 \\  
         & $K_{powerlaw}$        & $2.4 ^{+1.3}_{-0.20} \times 10^{-4}$ & $8.8^{+1.3}_{-4.4} \times 10^{-5}$ &  $< 7.0\times 10^{-5}$ \\  
\hline
 & $\chi^2_{\nu} (\nu)$             & 1.17 (11)     & 1.22 (10)     & 1.17 (9)              \\ 
\hline
 & Flux (10$^{-12}$erg~cm$^{-2}$s$^{-1}$) & 5.4     &  5.8          & 3.4             \\  
& Luminosity (10$^{32}$erg~s$^{-1}$) & 1.7     &  4.6   & 0.5        \\
 & Flux$_{cevmkl}$ (10$^{-12}$erg~cm$^{-2}$s$^{-1}$) & 0.66     &  2.3          & 3.4             \\  
& L$_{cevmkl}$ (10$^{32}$erg~s$^{-1}$) & 0.2   &  1.9    & 0.5        \\
 & Flux$_{power law}$ (10$^{-12}$erg~cm$^{-2}$s$^{-1}$) & 4.8     &  2.9          & $<$ 1.3             \\  
& L$_{power law}$ (10$^{32}$erg~s$^{-1}$) & 1.5   &  2.4    &  $<$ 0.2       \\
\enddata   \\
\flushleft{{\bf Notes.}
$N_H$ is the absorbing column, $\alpha$ is the index
of the power-law emissivity function ($dEM=(T/T_{max})^{\alpha -1} 
dT/T_{max}$), $T_{max}$ is the maximum temperature for the 
CEVMKL model, 
$K_{CEVMKL}$ is the normalization for the CEVMKL model, 
$K_{powerlaw}$ is the normalization for the power law model. 
${\bf \dagger}$ A 2$\sigma$ lower limit of the maximum plasma temperature is
stated because the parameter was not constrained in the higher bound limit,
the best fitting parameter is kT$_{max}$=49.8$^{<}_{-25.8}$.
Other than the above, all errors are given 
at 90\% confidence limit for a single parameter. 
The unabsorbed X-ray flux and the luminosities are given in the range 0.2-10.0 keV.
Solar abundances have been assumed. For luminosities distances are taken
from Table 1.} 
\end{deluxetable}

\clearpage

\begin{figure}[h] 
\begin{center}
\includegraphics[width=2.7in,height=4.5in,angle=-90]{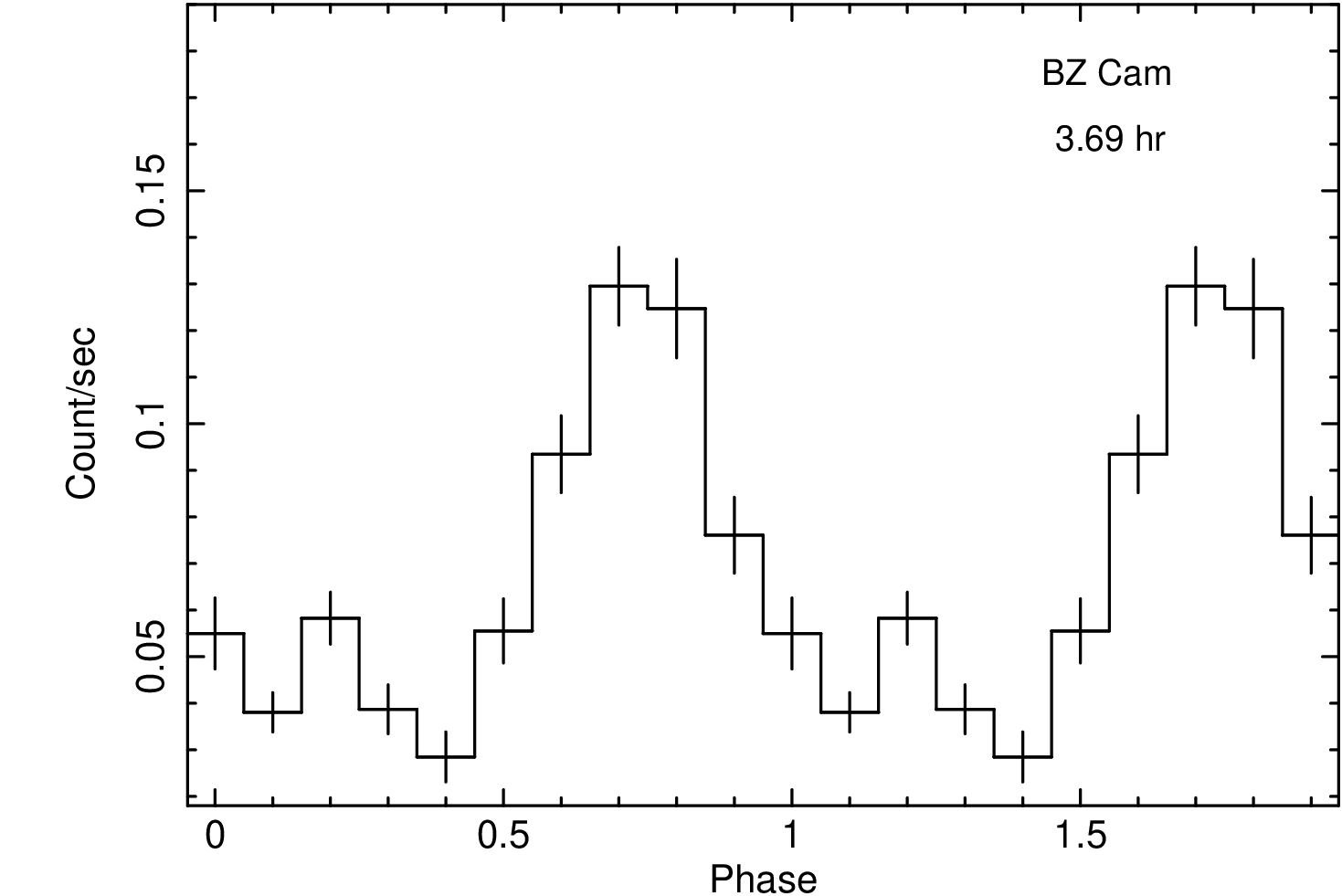}\\
\includegraphics[width=2.7in,height=4.5in,angle=-90]{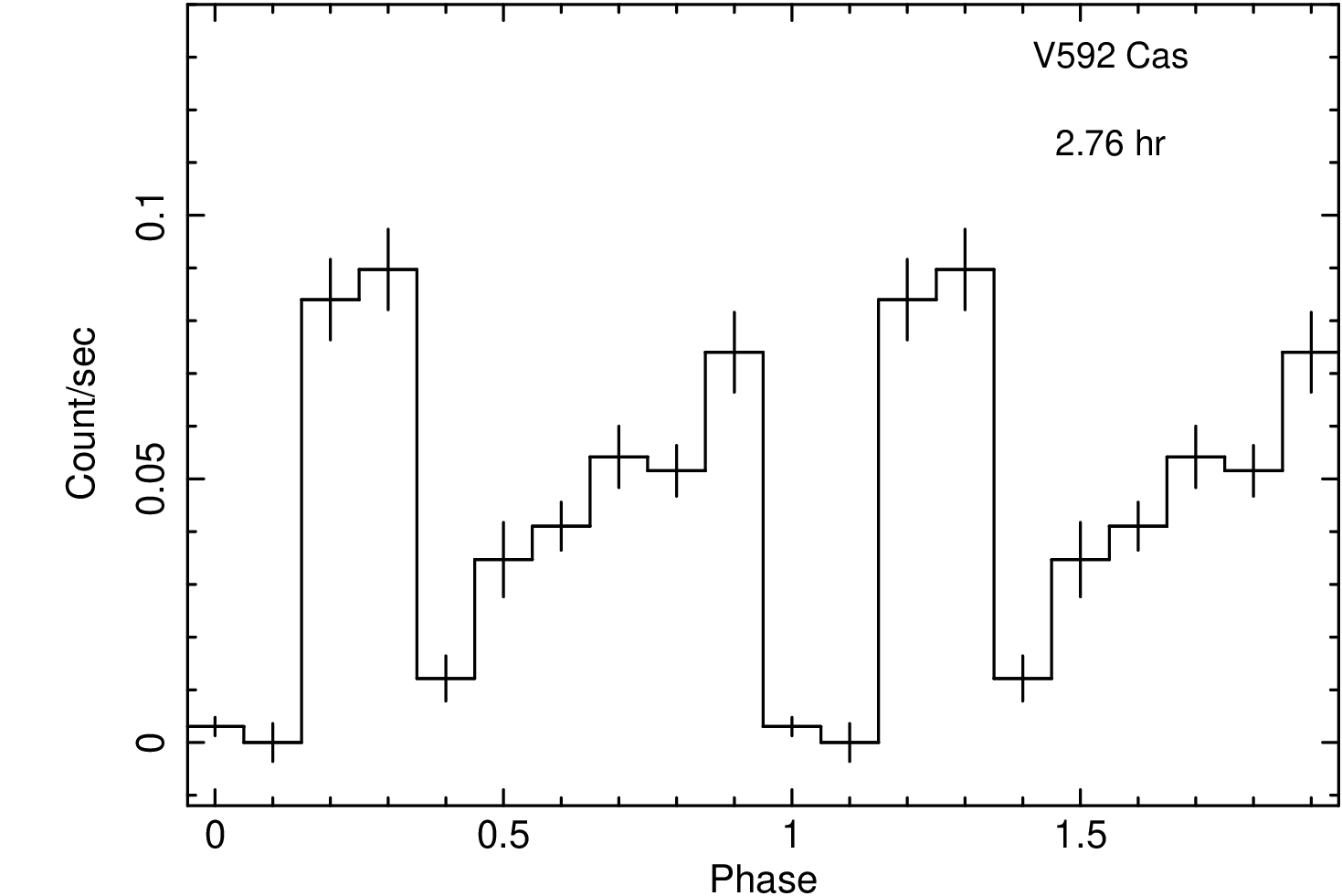}\\
\includegraphics[width=2.7in,height=4.5in,angle=-90]{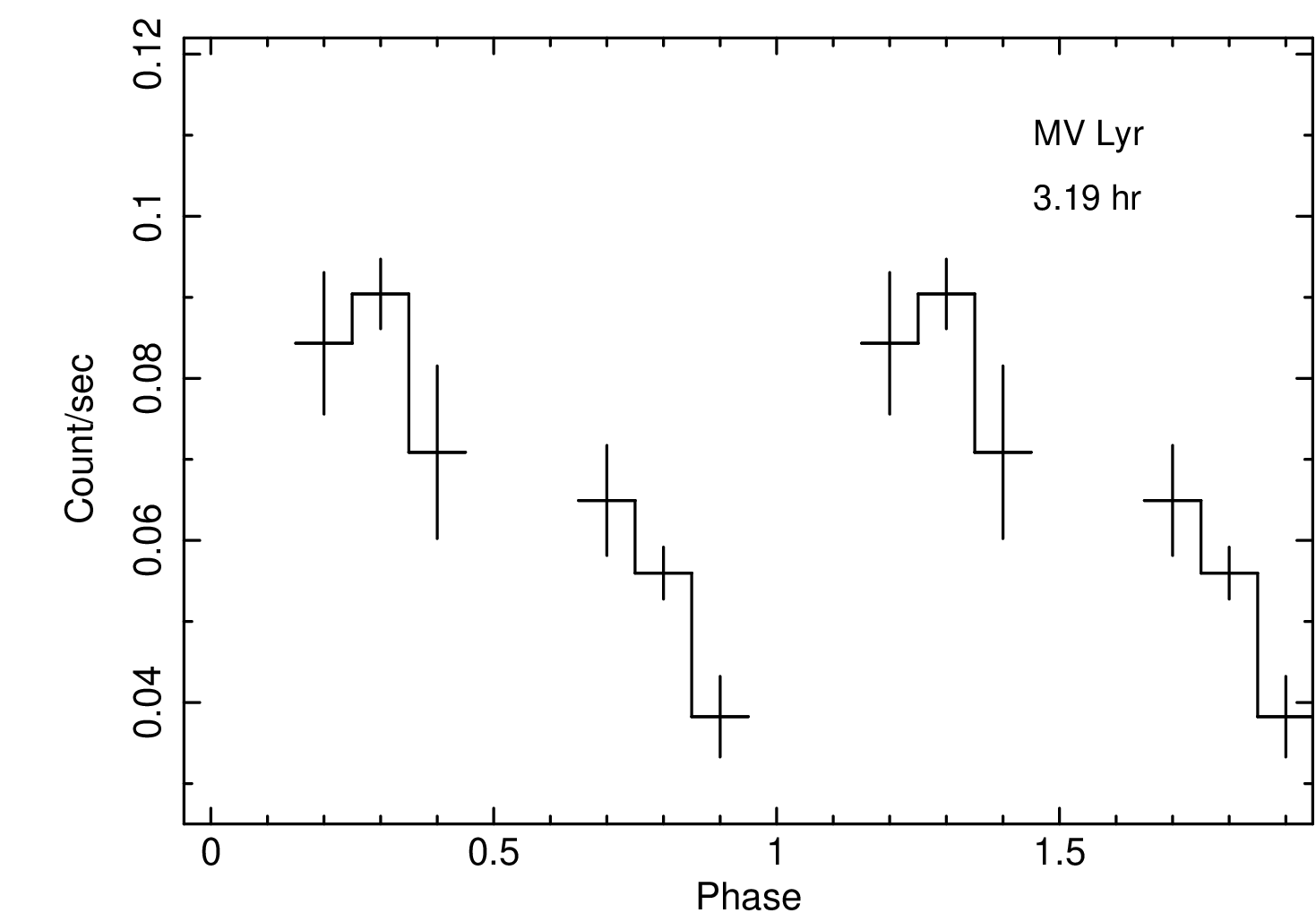}
\end{center}
\caption{
Mean \swi\ XRT light curves of BZ Cam, V592 Cas, and MV Lyr folded on their
orbital periods (see text for the Ephemerides), respectively. The orbital periods are labeled on the
panels. We caution that there is only about one-two cycle of orbital periods
for the systems and MV Lyr data does not have full orbital coverage. 
} 
\end{figure} 

\clearpage 

\begin{figure}[h] 
\vspace{2.cm} 
\includegraphics[width=4.9in,height=6.9in,angle=-90]{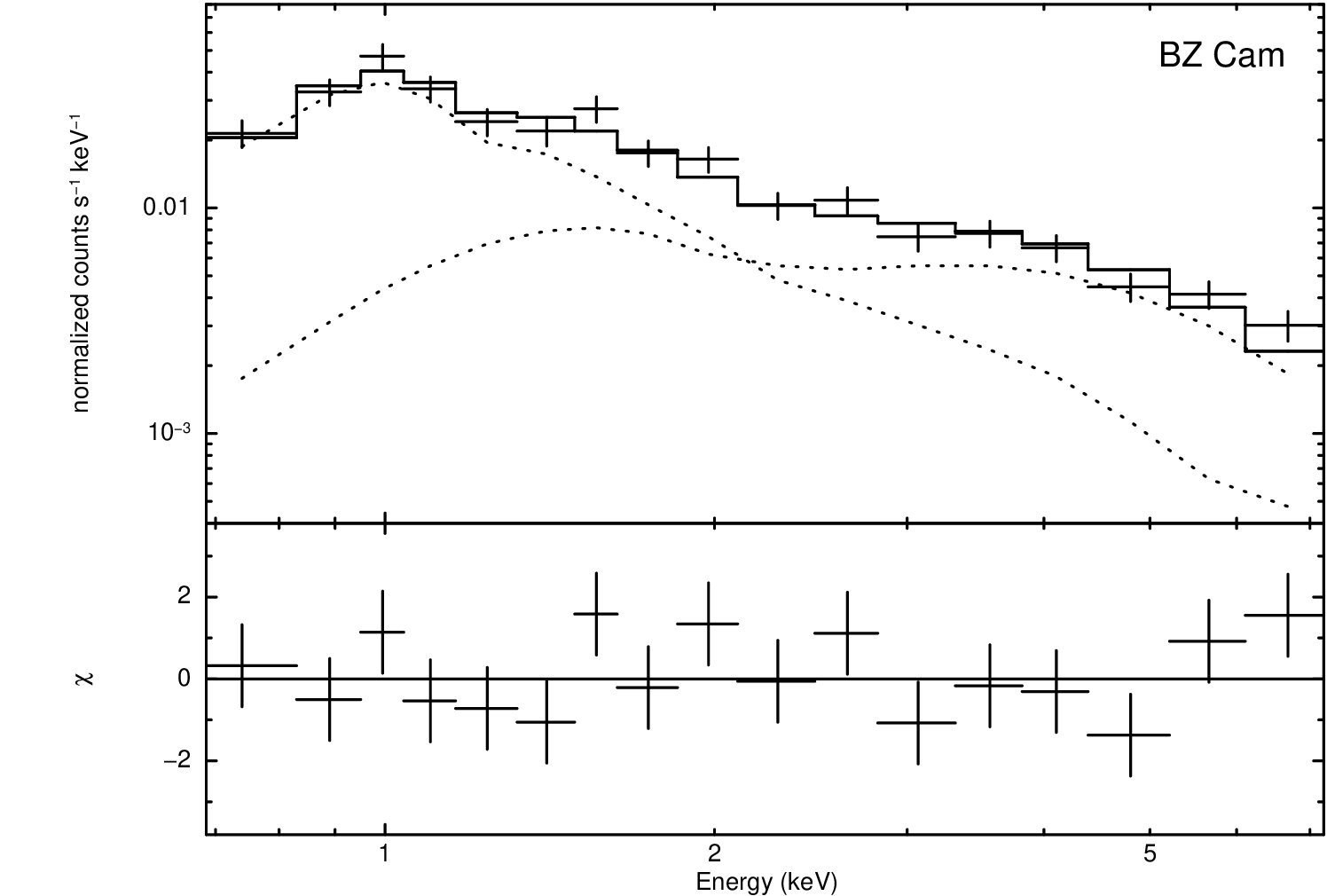}
\caption{
The \swi\ XRT spectrum of BZ Cam fitted with ($tbabs$*(CEVMKL+POWER)) model of
emission. The dotted lines show the contribution of the two fitted models.
The lower panel shows the residuals  in standard deviations (in sigma).}

\end{figure} 

\clearpage 

\begin{figure}[h] 
\vspace{2.cm} 
\includegraphics[width=4.9in,height=6.9in,angle=-90]{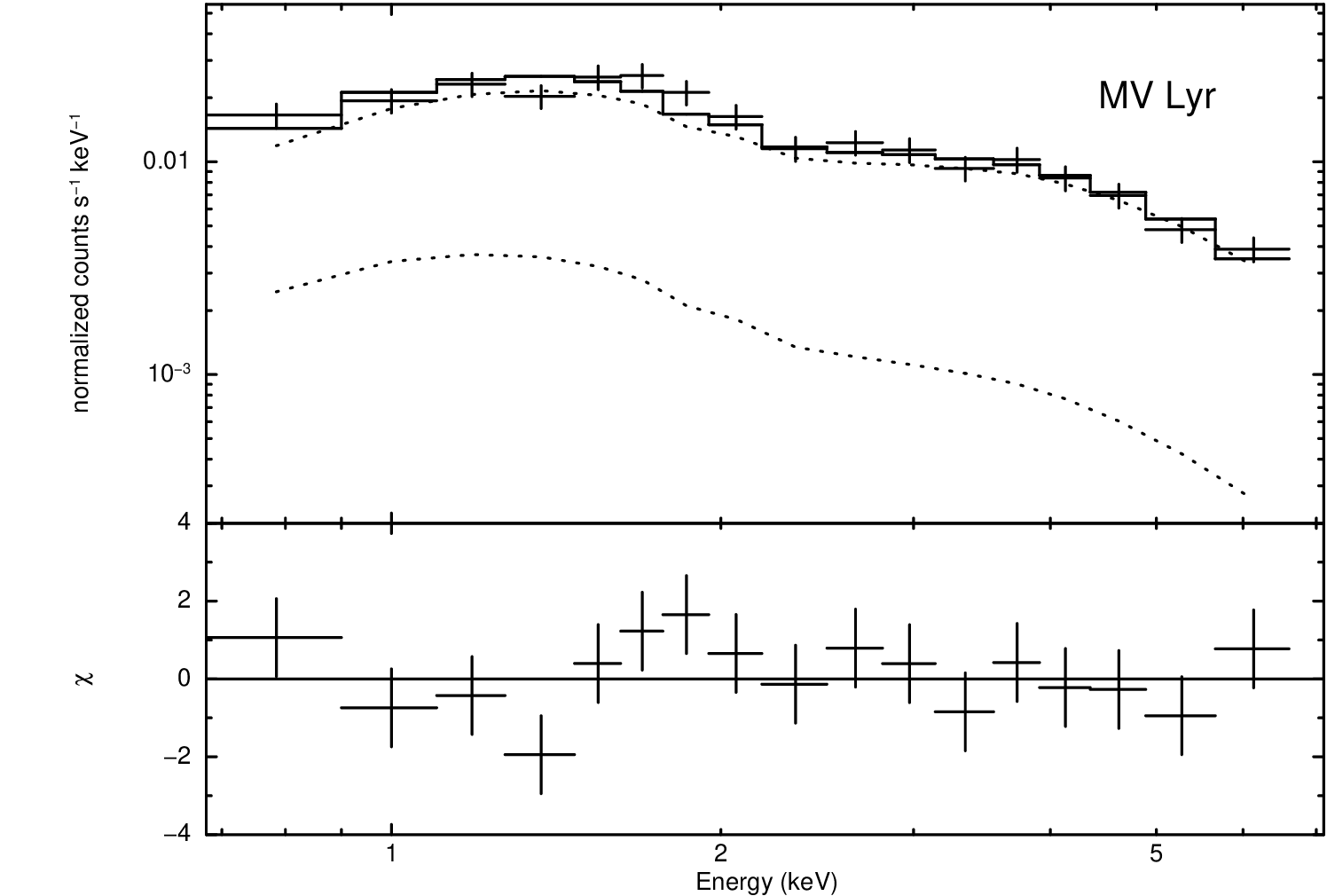}
\caption{
The \swi\ XRT spectrum of MV Lyr fitted with ($tbabs$*(CEVMKL+POWER)) model of
emission. The dotted lines show the contribution of the two fitted models.
The lower panel shows the residuals  in standard deviations (in sigma).}
\end{figure}

\clearpage

\begin{figure}[h] 
\vspace{2.cm} 
\includegraphics[width=4.9in,height=6.9in,angle=-90]{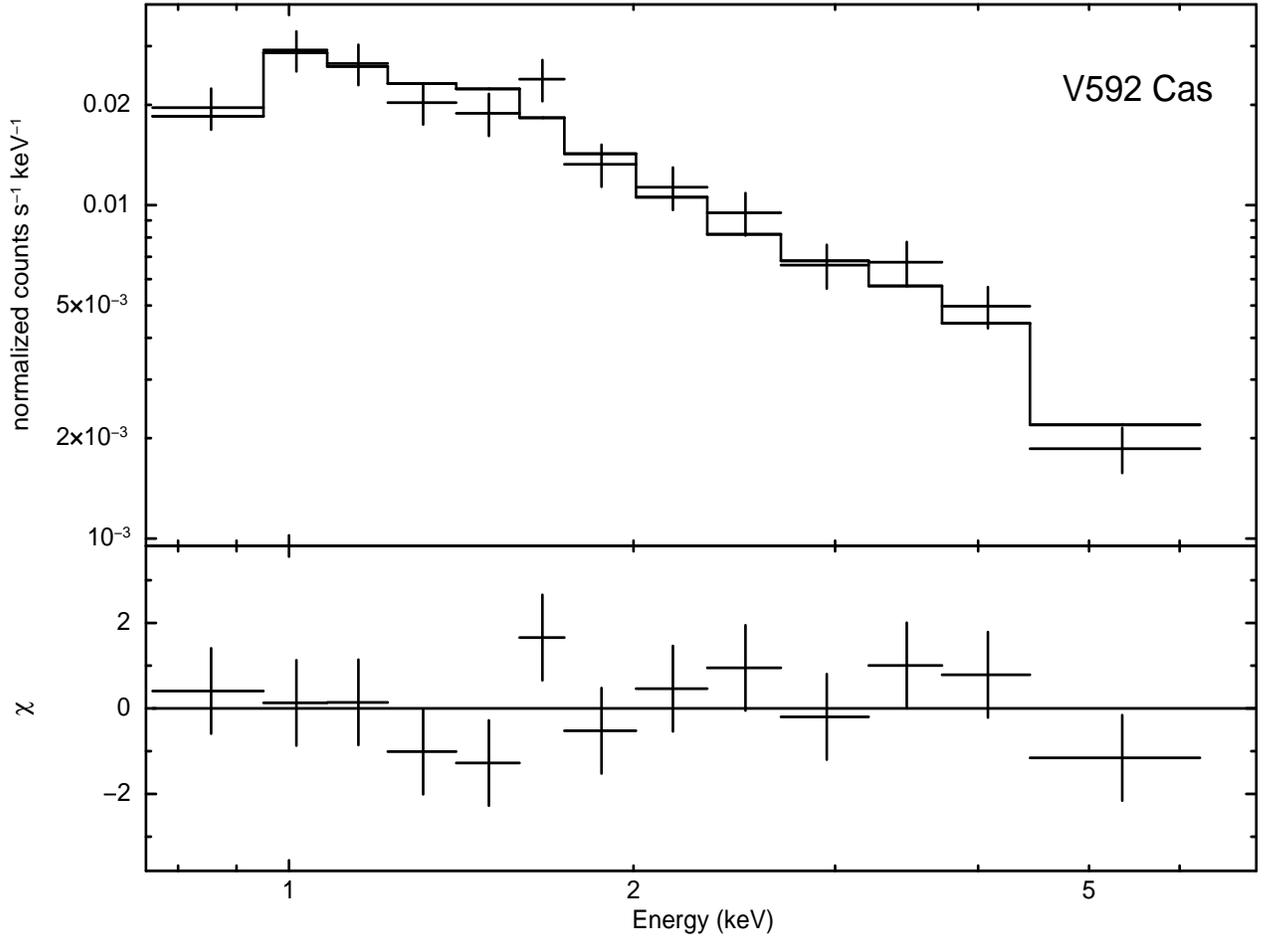}
\caption{
The \swi\ XRT spectrum of V592 Cas fitted with ($tbabs$*CEVMKL) model of
emission. 
The lower panel shows the residuals  in standard deviations (in sigma).
} 

\end{figure}

\end{document}